\documentclass[a4paper, 10pt]{article}
\usepackage{graphics}
\usepackage{amsmath}
\usepackage{amssymb}
\usepackage{geometry}
\usepackage[figuresright]{rotating}

\begin{document}

\title{DISSECTING INDIVIDUAL LIGAND-RECEPTOR BONDS WITH A LAMINAR FLOW CHAMBER}

\author{Anne Pierres, Joana Vitte, Anne-Marie Benoliel and Pierre Bongrand \\
Lab. Adhesion and Inflammation, UMR INSERM 600/CNRS 6212\\Marseille, France}

\maketitle

\begin{center}
{\bfseries  \itshape This un-edited manuscript was accepted for publication by Biophysical Reviews and Letters\\
and published in volume 1, pp 231-257 (2006)}
\end{center}

\begin{abstract}
The most important function of proteins may well be to bind to other biomolecules.
 It has long been felt that kinetic rates of bond formation and dissociation between
 soluble receptors and ligands might account for most features of the binding process.
 Only theoretical considerations allowed to predict the behaviour of surface-attached 
receptors from the properties of soluble forms. During the last decade, experimental
 progress essentially based on flow chambers, atomic force microscopes or biomembrane
 force probes allowed direct analysis of biomolecule interaction at the single bond level
 and gave new insight into previously ignored features such as bond mechanical properties or energy landscapes.
The aim of this review is (i) to describe the main advances brought by laminar flow
 chambers, including information on bond response to forces, multiplicity of binding
 states, kinetics of bond formation between attached structures, effect of molecular 
environment on receptor efficiency and behaviour of multivalent attachment, (ii)
 to compare results obtain by this and other techniques on a few well defined molecular
 systems, and (iii) to discuss the limitations of the flow chamber method.
  It is concluded that a new framework may be needed to account for the effective
 behaviour of biomolecule association.

{\it keywords : on-rate, off-rate, bond strength, intermediate state}
\end{abstract}

\section{Introduction : there is a need for a more complete description of molecular
interactions}
The main function of proteins may well be to bind to other proteins\cite{Creighton}. Also,
 it is a reasonable guess to assume that nearly half molecular species found on the membranes of 
living cells are involved in adhesive interactions\cite{Barclay}. Therefore, there is 
an obvious need for a theoretical framework to predict the outcome of encounters between
 two free or surface-bound ligand and receptor molecules.

A first quite naive way of fulfilling this task might consist of establishing a
 compendium of known ligands for each known molecule. Indeed, this primary task may
 certainly
 be considered as a useful step as exemplified by the success of the well known
 factsbook series, e.g.\cite{Isacke-Horton}. However, it has long been understood that
 such purely qualitative data were insufficient. For many years, the concept of
 affinity dominated most thinking about complex biological reactions\cite{Williams} and 
it might be felt that a good starting point for understanding the binding behaviour of
 biomolecules might consist of measuring the equilibrium affinity constants of association
 with their ligands. However, this simple view soon proved insufficient\cite{Williams}.
A prominent example is the adhesion of flowing leukocytes to surfaces coated with
ICAM-1, a ligand of leukocyte $\beta$2 integrins, or P-selectin, a ligand of the leukocyte
surface mucin PSGL-1. As was first shown by Lawrence and Springer in a well-known
 paper\cite{Lawrence-Springer}, ICAM-1 can bind to immobile leukocytes and keep them at rest
under flow, but it cannot stop rapidly moving leukocytes. Conversely, P-selectin
can dramatically slow down moving leukocytes and decrease their velocity by a factor of
nearly 100, but it cannot keep them immobile if the hydrodynamic flow is set up after a period
of rest. The difference between ICAM-1 and P-selectin could not be ascribed to a different
equilibrium constant of association with their ligands, but the authors suggested that
 this different behaviour could be
accounted for by a higher association and dissociation constant of bonds formed by
P-selectin, while affinities might be comparable. Another prominent example is the outcome
of interaction between T lymphocytes and antigen presenting cells expressing specific
complexes formed between histocompatibility molecules and peptides recognized by T cell receptors : it
is well known that this interaction may result in T lymphocyte full or partial activation, or
even paralysis, depending on the properties of interaction between T-cell receptors and
their ligand. It was clearly of utmost interest to relate lymphocyte behaviour to a 
quantitative interaction parameter. As recently summarized\cite{Ashwell}, many experiments
supported the conclusion that lymphocyte activation was determined by the lifetime of T-cell
receptor/ligand interaction rather than equilibrium constant.

Other experiments further challenged the simple view that kinetic constants, i.e. the
association on-rate and off-rate, might fully account for the association behaviour of
cells or molecules. Thus, it was righfully emphasized that a peculiarity of cell adhesion
receptors is that the bonds they form are often sujected to forces\cite{Puri}. Further, 
bond strength cannot be derived from kinetic paramers. Indeed,  
 chemical alteration of the CD34 ligand of leukocyte L-selectin
could selectively alter the resistance of L-selectin/CD34 bond to forces without changing
the lifetime of unstressed bonds. Other authors\cite{Pujades} reported that mutated
$\alpha 4\beta 1$ integrins bound fibronectin ligand with affinity and kinetic constants
comparable to those found on wild-type molecules, but cells expressing altered molecules
displayed impaired adhesion strengthening after initial attachment.

Also, it has long been emphasized that bond formation between two molecules is not an
all-or-none process, rather, it involves a number of intermediate binding states\cite
{Pierres1995,Pierres2002}. This multiplicity was recently illuminated with
the outstandingly resolutive biomembrane force probe\cite{Perret2004}, and as a consequence
it was well understood that bond rupture was dependent on bond
 history\cite{Marshall2005,Pincet2005}.

A final point is that there is no general way of deriving the rate of bond formation
between surface-bound molecules from the association constants measured on their soluble
forms. Indeed, the rate of bond formation between bound receptors is clearly dependent
on the distance between anchoring points and imposed orientation. This dependence is 
determined by molecular length and flexibility as well as molecular environment on the
membrane\cite{PierresCAC}. As a consequence of aforementioned data, it seemed warranted
to suggest\cite{PierresJIM} that molecular interactions should be described by at least
two {\bf functions} rather than {\bf parameters}, namely, the frequency of bond formation
as a function of distance between anchoring regions and the frequency of bond dissociation
as a function of applied force. In view of recent experiments disclosing the multiplicity
 of binding states, 
this simple view was insufficient to provide a satisfactory description of experimental
data yielded by presently available techniques.

Now, it would be a hopeless task to address experimentally aforementiond problems with
conventional techniques allowing only to monitor multiple interactions. Indeed, It is very
difficult to interpret results on the rupture of attachment between surfaces linked by
multiple bonds if the details of force sharing between bonds and possibility of rebinding
for detached molecules are not known\cite{Seifert2000}. Also, it would be difficult to 
interpret the association kinetics of receptor-bearing surfaces if molecular contacts
were not properly synchronized. It is therefore understandable that the 
development of methods allowing to study individual ligand-receptor bonds gave
a spectacular impetus to the dissection of molecular bonds.

Since the experimental study of ligand-receptor interaction heavily relies on the
development of efficient ways of probing single bond formation and dissociation, the aim of the
present review is twofold : first, we shall describe specific information obtained on
ligand-receptor interaction with a flow chamber method. When it is felt useful, we shall compare
these data to other pieces of information obtained with atomic force microscopy or biomembrane
force probes. Second, we shall discuss the present limitations of the flow chamber method and
pending issues concerning data interpretation.

\section{Experimental dissection of single ligand-receptor bonds \\ with a laminar flow
chamber}

\subsection{Basic principles.}

A schematic view of the laminar flow chamber is shown below.

\begin{figure}[h] \begin{center} \label{f1}
\includegraphics [width=10cm]{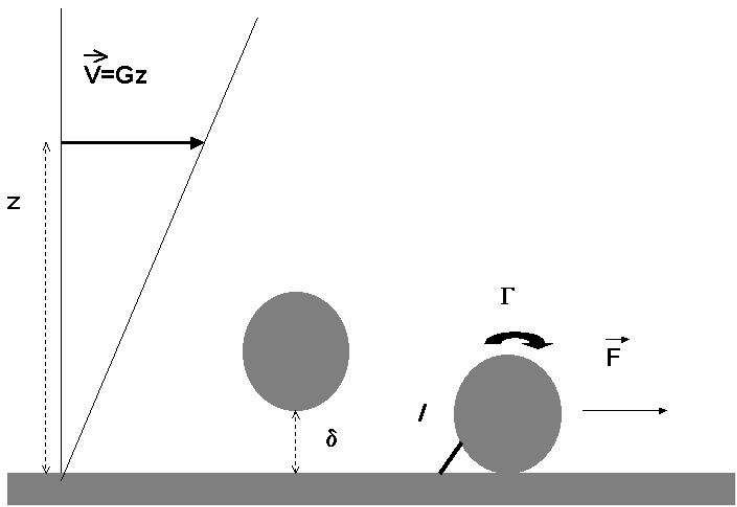}
\caption{\small A schematic view of the laminar flow chamber}
\end{center} \end{figure}

A receptor-bearing cell or particle is driven along a ligand-derivatized surface in a
laminar shear flow. Many investigators chose to operate under a wall shear rate G on 
the order of 100 s$^{-1}$ or more, thus mimicking hydrodynamic conditions encountered in blood
vessels\cite{Atherton-Born}. While this is certainly a reasonable choice to study the
properties of adhesion receptors mediating leukocyte-endothelium interaction, we found it
most rewarding to use 10-100 fold lower wall shear rates. The
advantage brought by this choice is twofold : first, the hydrodynamic drag is low enough
to allow a single molecular bond to maintain a particle at rest during a reasonable 
amount of time. Secondly, it is easy to monitor freely moving particles. This allows
quantitative study of the frequency of bond formation, which is a currently important issue.

\subsubsection{Kinetics}
Data interpretation requires a thorough understanding of the motion of a free cell-like object
near a wall in shear flow. The motion of a neutrally buoyant sphere near a plane in laminar
shear flow was solved numerically and results were nicely summarized in an often-quoted 
paper\cite{Goldman}. The main conclusions are as follows : when the distance $z$ between the
sphere center and the wall is much higher than the sphere radius, the velocity 
$U$ is about $Gz$, where $G$ is
the wall shear rate. However, when the sphere comes close to the surface, its velocity 
is expected to vanish very slowly. As a consequence, measuring the velocity may allow a 
very accurate estimate of the sphere-to-surface distance. We found it convenient to use
the following polynomial approximation of Goldman's data\cite{Pierres2001} :

\begin{equation}
U/aG = \exp(0.0037644 \ln(\delta/a)^3 + 0.072332 \ln(\delta/a)^2 + 0.54756 \ln(\delta/a)
 + 0.68902) 
\label{eq1}
\end{equation}

where $\delta$ is the distance between the sphere and the wall and $a$ is the sphere radius. Two
comments may be useful.

First, it may be noticed that $U/aG$ is close to 0.7, 0.5 and 0.4 respectively when $\delta/a$ is
about 0.1, 0.01, 0.001, corresponding to a distance of 250 nm, 25 nm and 2.5 nm for a cell-size
sphere of 5 micrometer diameter. Obviously, the significance of the lowest values may be
questioned, since cell surfaces are not smooth at the nanometer level.
 The relevance of Eq.~(\ref{eq1}) to actual cells was 
checked experimentally\cite{Tissot92}~: Goldman's theory was concluded to apply to cells provided
a phenomenological friction coefficient was added (this was estimated at 74 nanonewton.s/m), and
the cell-to-surface distance was about 1.5 micrometer. This gap might be partly due to 
cell surface protrusions, but other experiments supported the importance of short range repulsion
by components of the pericellular cell matrix or glycocalyx\cite{Tissot91}

Second, the relevance of this theory to microspheres of 2.8 µm diameter and 1.3 g/l density 
was studied exerimentally\cite{Pierres2001}. Full agreement was obtained provided brownian
motion was included. This required to take care of hydrodynamic interaction between spheres
and surfaces. This interaction resulted in increasing the effective friction cofficient for
 displacements parallel to the wall by the following dimensionless factor~:

\begin{equation}
F_x = \exp(0.00332 \ln(\delta/a)^3 + 0.019251 \ln(\delta/a)^2 - 0.18271 \ln(\delta/a)
+0.32747)
\label{eq2}
\end{equation}

Thus, due to vertical Brownian motion, the average distance between spheres and the surface was
about 120 nm (in accordance with Boltzmann's law) and hydrodynamic forces reduced the diffusion
coefficient parallel to the wall by about 50 \%. The interest of using these microspheres instead
of cells was that i) surface structure could be better controlled, and ii) computer-assisted image
analysis allowed accurate determination of position with better than 50 nm resolution.

The results presented above strongly suggest that the distance between a flowing cell or 
microsphere and a ligand-coated surface may be higher than the length of most adhesion molecules.
Indeed, common receptors such as integrins are about 20 nm length, and especially large
adhesion molecules such as P-selectin (CD62L) or its ligand PSGL-1 (CD62L) are about 40 nm
long. Bond formation thus requires a transient decrease of particle-to-surface distance.
It is therefore important to consider the details of vertical motion to estimate the rate of 
surface molecule associations. This vertical motion may be determined by the 
structure of glycocalyx elements or surface rugosity when cells are studied.
 When microspheres are considered,
the hydrodynamic interaction between the sphere and the wall may reduce diffusion
perpendicular to the wall by a factor $F_z$ close to $a/\delta$ at low sphere-to-surface
distance. The following approximation was found convenient\cite{Pierres2001}~:

\begin{equation}
F_z = \exp(0.0057685 \ln(\delta/a)^3 + 0.092235 \ln(\delta/a)^2 - 0.52669 \ln(\delta/a)
+ 0.76952)
\label{eq3}
\end{equation}

Another point deserving some emphasis is the duration of particle-surface contact. A sphere 
flowing close to the surface is expected to rotate with angular velocity (in radian/s) 
$0.54 U/a$. The relative velocity between the particle surface and the wall is thus 0.46 $U$.
If the length of adhesion receptors is of order of 20 nm and particle velocity is of order of
10 µm/s, molecular encounter duration is therefore of order of 4 ms.

\subsubsection{Mechanics.}
As shown on Fig.~\ref{f1}, a particle deposited on the chamber floor is subjected to a
drag force $\vec{F}$ and torque $\vec{\Gamma}$. When the particle is spherical, these 
are given by the following equations\cite{Goldman}~:
\begin{equation}
\vec{F} = 32.05 \mu a^2 G
\label{eq4} \end{equation}

\begin{equation}
\vec{\Gamma} = 11.86 \mu a^3 G
\label{eq5} \end{equation}

where $\mu$ is the medium viscosity. As previously reported\cite{Pierres1995}
, the tensile force T exerted on the bond is given by~:
\begin{equation}
T \approx (F + \Gamma /a)\sqrt{a/2L}
\label{eq6} \end{equation}

As a rule of thumb, in a medium of 0.001 Pa.s viscosity such as water,
 the force (in piconewton)
exerted on a bond of 20 nm length maintaining at rest a sphere of 1.4 micrometer radius
in presence of a wall shear rate $G s^{-1}$ is about 0.5~$G$. As shown on Eq~\ref{eq6}, the 
dependence of the force on bond length is fairly weak. When particles are cells, 
the force on the bond may be much lower due to a higher lever arm. This lever arm
may be of order of the sphere radius\cite{Alon95} or even severalfold higher due
to the formation of long tethers\cite{Diamond2000}. In this case, the force is
close to the value given in Eq.~\ref{eq4}.

\subsection{Detection of single bonds - Lifetime determination} 

The first experimental observation of single bond rupture was probably performed in
Harry Goldsmith's laboratory\cite{Tha86}. Osmotically sphered erythrocytes were coated with
minimal amounts of antibodies and subjected to laminar shear flow in a moving tube allowing
cells to remain  a substantial amount of time in the field of a fixed microscope.
 Hydrodynamic forces made erythrocytes 
encounter each other, then form doublets that rotated, and experienced a growing distractive
 forces until they were ruptured. 
This allowed the first determination of so called {\it unbinding forces}, leading to the
estimate that the mechanical strength of an antigen-antibody bond was close to 24 pN. However,
individual encounters could not be monitored for a sufficient period of time
 to allow systematic monitoring of bond formation and 
dissociation.

Only a few years later did a laminar flow chamber allow extensive observation of blood
leukocytes flowing along a surface coated with endothelial cells\cite{Kaplanski93}. When subjected
to a hydrodynamic force of a few piconewtons, cells exhibited multiple transient arrests
of widely varying duration. These arrests could be ascribed to well defined interactions
 between endothelial E-selectin and leukocyte ligand since they were inhibited by specific
antibodies. In view of previous theoretical\cite{Bell78} and 
experimental\cite{Tha86,Evans91} work, it was assumed that a single molecular bond could
 mediate cell arrest. Analysing the statistics of arrest duration, it was concluded that the
dissociation rate of an E-selectin mediated bond was about 0.5 s$^{-1}$.

Now, there remained to prove that this approach indeed allowed detection of single bonds, which is
by no means trivial\cite{Zhu2002}.

A reasonable test might consist of performing limiting dilution analysis and checking that
arrest frequency was proportional to the first power of ligand or receptor surface density. This was
indeed performed by several investigators\cite{Pierres1995,Alon95,Pierres1996}. However,
 this cannot be
considered as a formal proof that single bonds are actually detected. Indeed, if only double bonds 
were detected and receptor preparation contained a limited fraction of aggregates required for
arrest, a linear relationship would be expected to occur between receptor density and arrest
frequency. Other arguments might be obtained by measuring bond frequency and controlling surfaces
for molecular distribution. Also, if it is accepted that the wall shear rate is low enough
to allow single bond detection, the shortest binding events may reasonably be ascribed to 
single bonds.

Now, if binding events were mediated by single bonds with a dissociation rate $k_{off}$, the number
of cells or particles remaining bound at time t after initial arrest should be given by the simple
relationship :

\begin{equation}
N(t) = N(0) \exp{-k_{off} t}
\label{eq7} \end{equation}

However, while this simple relationship was indeed found to hold under some experimental 
situations\cite{Alon95,MassonGadais99}, it was soon emphasized that

{\it i)} Detachment curves obtained with different ligand-receptor models were not
 linear\cite{PierresLivrePB94,Pierres1995}.

{\it ii)} This nonlinearity could be accounted for by different mechanisms such
as formation of additional bonds after arrest\cite{Pierres1996} or time-dependent strenghtening
of newly formed bonds as a consequence of the occurrence of several energy barriers along the
reaction path\cite{Pierres1995}.

{\it iii)} Conversely, the occurrence of a linear relationship 
between $\ln{N(t)}$ and time could not be considered as a formal proof that single bond lifetimes
 were indeed measured, if time resolution or the range of measured durations were not
 sufficient\cite{CellMechanics94}.

\begin{figure} [h] \begin{center}
\includegraphics[width=10cm]{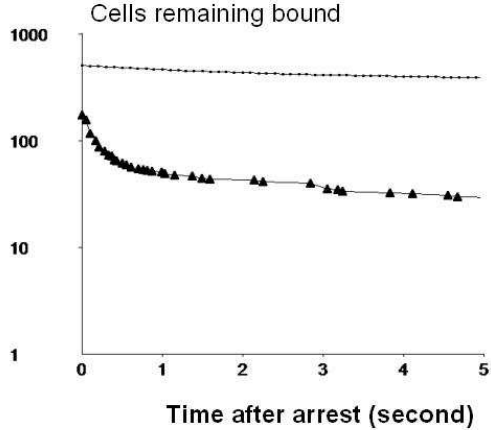}
\vspace*{8pt}
\caption{\small Two typical detachment curves. The detachment plot determined with
keratinocytes moving along collagen-coated surfaces (dots)
was a straight line, in contrast with the curve determined on monocytic
THP-1 cells moving along fibronectin-coated surfaces (triangles) \label{f2}}
\end{center} \end{figure}

Typical detachment curves are shown on Fig.~\ref{f2}. A common procedure consisted of 
determining dissociation rates by considering only the initial linear part of unbinding
plots\cite{Puri}.

As shown on Table ~1, when different ligand-receptor couples were studied, measured 
bond lifetimes spanned at least a 1000 fold range. Clearly, it would be of high interest to
relate experimental bond lifetimes to functional properties of adhesive interaction they 
mediate, such as reversibility.

Another interesting point is that bond  lifetime modulation was found to be used by living cells as
a mechanism of controlling adhesion. Thus, integrins constitute an important family of 
adhesion receptors that are subjected to extensive regulation. The involved mechanisms 
were studied by many authors, whose studies revealed the functional importance of 
affinity changes due to a passage between different conformational states. It was therefore
important to assay the importance of these changes with a flow chamber. Integrin-mediated
interaction between collagen surfaces and human keratinocytes was studied : when cells
were treated with monoclonal antibodies known to enhance or decrease integrin activity, the 
bond lifetime was found to vary accordingly : function-enchancing antibodies triggered 
sixfold increase of bond lifetime, and function inhibiting antibodies triggered up to 
13fold decrease of bond lifetime\cite{MassonGadais99}.

\subsection{Studying the force dependence of bond lifetime}

\subsubsection{Use of the laminar flow chamber to determine force dependence of bond lifetime}

The interest of investigators in the force dependence of bonds was strongly stimulated by
a theoretical report from George Bell\cite{Bell78} who elaborated a framework
to address interactions between cell surface receptors. He suggested that the dissociation
rate $k_{off}(F)$ of a bond subjected to a disruptive force $F$ might be calculated with
 the following equation~:

\begin{equation}
k_{off}(F) = k_{off}(0) \exp(Fd/k_BT)
\label{eq8} \end{equation}

where $T$ is the absolute temperature, $k_B$ is Boltzmann's constant, and $d$ should represent
the range of intermolecular forces. 
Eq.~(\ref{eq8}) is now referred to as Bell's law. It might be somewhat justified on 
theoretical grounds, starting from the early Eyring and Kramer theories of reaction
rate\cite{Evans97}. Now, considering antigen-antibody bonds, Bell estimated $d$ at about
 0.5 nm,a value somewhat intermediate between the depth of an antibody cleft and the range of individual
interactions. The force coefficient $f_0$ = $k_BT/d$ might thus be expected to be of order of
8 pN.

The first in-depth experimental check of Bell's idea was performed  
by Alon and colleagues\cite{Alon95}. These authors used a laminar flow chamber to monitor the
motion of blood neutrophils driven along surfaces coated with various surfaces densities of 
P-selectin molecules : the wall shear rate ranged between about 17 and 73 $s^{-1}$. Chosen
values were thus substantially lower that those previously used to mimick blood 
flow\cite{Lawrence-Springer}. When the surface density of P-selectin was higher than about
30 sites per $\mu m^2$, close to the physiological range found on activated endothelial cells,
cells interacting with the flow chamber exhibited a jerky motion closely resembling the 
rolling phenomenon. However, when the surface density of P-selectin was lowered down to
1 site per $\mu m^2$, cells moving close to the surface displayed periods of displacement
with constant velocity interspersed with transient arrests of varying duration, closely similar
to the motion observed with lower wall shear rates and higher ligand densities\cite{Kaplanski93}.
These arrests followed first order dissociation kinetics, and the dissociation rate increased
when the shear rate was increased. The authors concluded that they analyzed single molecule
binding events, and that the dissociation rate followed Bell's law. They were able to derive
the force from calculated hydrodynamic forces exerted on cell by a simple geometrical model
of cell shape. The zero-force dissociation rate was 0.95 $s^1$, in close accordance with 
the estimate of E-selectin bond lifetime\cite{Kaplanski93}, and the characteristic distance
 parameter $d$ of
Eq.~\ref{eq8} was 0.049 nm, which was much lower than Bell's estimate. This value was 
considered to be consistent with those found for hydrogen or calcium coordination bonds.

During the next years, many authors reported estimates of zero-force dissociation rates and
force parameters $k_B T/d$ for different ligand-receptor couples. Some results are summarized
on Table~1. For convenience, rows were ordered according to zero-force dissociation rate.

\begin{center}
{\bfseries Table~1 Dissociation parameters measured on different ligand-receptor couples.}
\begin{tabular}{lccc} \hline
Ligand-receptor couple & $k_{off}(0)$ & $k_BT/d$ & Reference\\
& $s^{-1}$ & piconewton &\\ \hline
CD2/CD48 & 7.8 & 32 &\cite{Pierres1996}\\
L-Selectin/CD34 & 6.6 & 218 & \cite{Alon97}\\
P-Selectin/PSGL1 & 0.95 & 82 &\cite{Alon95}\\
IgG/anti-IgG (transient state)& 0.9 & 53 & \cite{Pierres1995}\\
E-Selectin/PSGL1 ? & 0.70 & 138 & \cite{Alon97}\\
E-SelectinPSGL-1 ? & 0.50 & - & \cite{Kaplanski93}\\
E-Cadherin/E-Cadherin & 0.45 & 2.4 & \cite{Perret02}\\
Antibody/Group B antigen & 0.04 & 35 & \cite{Tees96}\\
VLA2/Collagen & 0.01 - 0.2 & - & \cite{MassonGadais99}\\
IgG/Protein G & 0.006 & 11 & \cite{Kwong96}\\
\hline   \end{tabular}
\end{center}

Clearly, zero-force dissociation rates displayed a wide range (a thousandfold) of 
values, and there was no obvious correlation between $k_{off}$ and $k_Bt/d$.

Now, while Bell's model was considered to account satisfactorily for experimental
results found in many studies\cite{Chen01}, additional complexity was soon disclosed.

An important point is the recent experimental demonstration of the existence of
 so-called {\it catch bonds}\cite{Marshall03}
that had been fancied in the late eighties by Dembo and colleagues\cite{Dembo88}. These 
authors had emphasized that thermodynamics only imposed that the affinity of a given 
bond be decreased by a disruptive force, but association and dissociation rates (the
ratio of which gives the affinity constant) might in principle by either decreased or 
increased by forces. Thus, they suggested to call {\it slip bonds} the molecular bonds
that displayed decreased lifetime in presence of disruptive forces, as intuitively 
expected, and {\it catch bonds} the bonds that exhibited opposite behavior. It was only recently
that Zhu and colleagues were able to convince the scientific community that catch bonds
indeed existed, with both a laminar flow chamber and atomic force microscopy. The 
conclusion was that selectin-mediated bonds might display increased lifetime in presence
of a low disruptive force.

\subsubsection{Analysis of bond rupture with other methods : atomic force microscopy and
biomembrane force probe}

In a flow chamber, a bond is subjected to a {\it constant force}, and measuring bond lifetime
makes it intuitively obvious that bond rupture is a stochastic phenomenon that can only
be accounted for by a rupture probability.

Now, when molecular attachments were subjected to a rapidly {\it increasing} force with an atomic
force microscope\cite{Florin94}, it appeared intuitively obvious that the relevant parameter
was the {\it unbinding force}. It seems that the authors felt they were probing the low temperature
limit of bond rupture and they were measuring an intrinsic and time-independent parameter. It was
only a few years later that the maximum force resisted by a bond subjected to increasing
effort was shown to be dependent on the rate of force increase,
 i.e. the {\it loading rate}\cite{Merkel99}, as expected from Bell's law. Only when the loading
rate was varied it became possible to achieve experimental determination of both zero-force
dissociation rate and aforementioned parameter $k_BT/d$. However, it immediately appeared that
single bonds behaved in a less simple way than initially expected, and bond rupture appeared
to involve the sequential crossing of several energy barrier. Each crossing appeared 
to follow Bell's law. Thus, experimental results obtained with different methods cannot be
compared before we address this supplementary level of complexity.

\subsection{Analyzing the complexity of ligand-receptor association}

\subsubsection{Dissection of single molecular bonds with a flow chamber}

For many years, it seemed acceptable to view ligand-receptor association as an all-or-none
phenomenon : molecules should be either bound or separated, but intermediate complexes 
were considered as so transient that they should not be detectable unless very 
rapid kinetic methods were used.

However, when single molecular bonds began being studied, it soon appeared that association
indeed involved readily detectable intermediate states. Indeed, when microspheres coated with
 anti-rabbit immunoglobulin antibodies were made to interact under low shear with surfaces
coated with rabbit immunoglobulins, they displayed binding events with a short lifetime 
of order of 1 second that was fairly insensitive to shear. However, a significant proportion
of binding events lasted several seconds or even tens of seconds or more\cite{Pierres1995}.
 It was concluded that 
antibody binding was a multiphasic process involving the formation of a transient intermediate
state and further strengthening.

Similar conclusions were obtained when streptavidin-coated spheres were made to interact
with biotinylated surfaces\cite{Pierres2002}. Transient binding states were observed under
conditions where the forces imposed on bonds varied between 3.5 and 11 pN. Since this lifetime
was much shorter than expected in view of Bell's law and previous experiments done with
atomic force microscopy\cite{Florin94}, it was concluded that these results revealed the
existence of intermediate binding states. Also, since a modest increase of the shear rate
 and hydrodynamic forces
resulted in dramatic decrease of the frequency of binding events without any substantial change
of other kinetic parameters, it was suggested that experimental findings might reveal the 
existence of a still shorter binding state that might be highly sensitive to forces.
Finally, intermediate binding states with a similar lifetime of order of one second were
reported between fibronectin-coated surfaces and monocytic cells expressing integrin receptors
for fibronectin\cite{Vitte04}.

\subsubsection{Dissecting individual bonds with atomic force microscopy or the biomembrane
force probe}
As explained above, when the loading rate was systematically varied, first with the biomembrane
force probe\cite{Merkel99} and then with atomic force 
microscopy\cite{Zhang02,Hanley03,Panorchan06}, rate and force constants could be obtained, and
it is interesting to compare data obtained on similar molecular systems with different techniques. 
Selected results obtained on the homotypic interaction between the outer two domains of
E-cadherin are shown on Table~2.

\begin{center}
{\bfseries Table~2. Rupture of  homophilic associations \\between E-cadherin external domains EC1-EC2\\}
\begin{tabular} {lccc} \hline
Method & $k_{off}(0)$ & $k_BT/d$ & Reference\\
 & $s^{-1}$ & $pN$ & \\ \hline
Flow Chamber & 0.45 & 2.4 & \cite{Perret02}\\ \\
Biomembrane force probe & 0.5 - 1 & 6 - 7 & \cite{Perret04} \\
 & 8 - 11 & 6 - 7 & \cite{Perret04} \\
\hline \end{tabular} 
\end{center}

It is concluded that results obtained with the biomembrane force probe and the flow chamber
are consistent. The former method displayed additional capacity to reveal a supplementary 
barrier.

Now, results obtained on the P-selectin model are displayed on Table~3 :

\begin{center}
{\bfseries Table~3. Rupture of individual bonds\\ formed between P-selectin and PSGL-1 ligand\\}
\begin{tabular} {lccc} \hline
Method & $k_{off}(0)$ & $k_BT/d $ & Reference\\
 & $s^{-1}$ & $pN$ & \\ \hline
Flow Chamber & 0.95  & 82  & \cite{Alon95}\\ \\
Atomic Force Microscopy & 0.2 & 30 & \cite{Hanley03} \\ \\
Biomembrane Force Probe & 0.37 & 18 & \cite{Evans04} \\
 & 8-12 & 6\\
\hline \end{tabular} 
\end{center}

Here, the flow chamber yielded much higher force constant than the other two techniques.
It must be pointed out that the wall shear rate used in this case was much higher than
in \cite{Perret02}. Under these conditions, binding events corresponding to the parameters
obtained with the other two methods might have been difficult to detect.

Thus, experimental data obtained with the flow chamber, atomic force microscopy and the 
biomembrane force probe convincingly demonstrated that bond formation and dissociation 
involve the formation of a hierarchy of binding states of widely varying duration, reflecting
the existence of a sequence of barriers in the association or dissociation path, viewed as 
a valley in the energy landscape of the ligand-receptor molecular complex\cite{Eyring35}.
Probably the laminar flow chamber is most effective in revealing the properties of the 
external part of the energy landscape\cite{Pierres2002}, since it allows to probe
bonds within a millisecond after formation, subjecting them to forces as low as a few 
piconewtons.

\subsection{Experimental study of the rate of bond formation between surface-attached molecules}
\subsubsection{Results obtained with flow chambers}

Determining the rate of bond formation between single molecules is by no means an easy task.
A typical experiment would consist of maintaining molecules within binding range for a given
amount of time, then determining whether binding occurred. There are three main problems to
solve :

First, how can we define two molecules as bound when there is a number of transient binding states,
and there is a need to chose an arbitrary threshold duration and/or resistance to force allowing 
to define two molecules as bound.

Secondly, unless molecules are freely diffusing in a given region of space,the frequency of
bond formation is highly dependent on molecular motion, and particularly the motion of binding
sites.

Thirdly, if experiments are done with a number of molecules, initial states must be known
with higher accuracy. Indeed, if you bring together two surfaces coated with ligand and receptors
with low binding probability, it is usually accepted that detected binding events essentially
represent single bonds, based on Poisson law. Thus, there is no need to know the initial 
state of all potential binders. On the contrary, an experimental rate of bond formation is 
meaningless if the initial state of receptors and ligands is not known.

As a consequence, when laminar flow chambers were first used with low shear rate to monitor
bond formation between blood neutrophils and activated endothelial cells\cite{Kaplanski93},
the frequency of bond formation was readily determined by monitoring the motion of cells
moving in contact with the substrate. However, the experimental frequency
 (0.04 $s^{-1}$) could not be used to derive intrinsic molecular parameters since the 
density and topography of binding molecules were not known.

In a later study\cite{Pierres97}, a better defined model system was used to study the 
kinetics of bond formation between surface-attached molecules. Microspheres of 2.8 
micrometer diameter were coated with CD48 molecules using antibodies as linkers. They
were then driven along surfaces derivatized with CD2, a CD48 ligand, under low shear.
Numerous binding events were recorded, and arrest frequency was determined as a function
of sphere velocity immediately before arrest. Ignoring brownian motion, it was thus 
possible to relate the sphere velocity to distance from the surface. This allowed 
 derivation of a tentative relationship between the binding frequency $k(z)$ of two molecules
CD2 and CD48 maintained at distance $z$, and the binding rate $P(z)$ of a sphere located
ad distance z from the surface. Denoting the binder surface density on beads and 
surfaces as $\sigma _b$ and $\sigma _s$ respectively :

\begin{equation}
k(z) = \frac {d^2P/dz^2}{4\pi ^2 a z \sigma _b \sigma _s}
\label{eq9} \end{equation} 

The binding frequency was found to be inversely related to the cube of the intermolecular 
distance, and it was estimated at 0.03 $s^{-1}$ at 10 nm separation.

Since the above conclusions relied on unwarranted neglect of Brownian motion, a refined model was 
used to obtain a more accurate derivation of molecular association rates\cite{Pierres98}.

Microspheres were coated with recombinant outer domains 1 and 2 
(denoted as EC1-2) of cadherin 11, a
homotypic adhesion molecule. They were made to interact with molecularly smooth
mica surfaces that were coated with EC1-2. In order to ensure proper orientation
of bound molecules, adsorption was obtained through hexahistidine moieties that
were added to the tail of EC1-2. Multiple trajectories
were monitored, and a total of nearly one million positions were recorded.
First, acceleration curves were built by plotting the average particle
acceleration versus velocity. These experimental plots were fitted with
curves obtained by computer simulations that accounted for i) Brownian motion
parallel and perpendicular to the wall, ii) hydrodynamic interaction between
particle and the wall, and iii) van der Waals attraction
between spheres and the wall, as accounted for by an adjustable Hamaker constant.
Electrostatic interactions were neglected since experiments were performed
in media of high (physiological) ionic concentration. Fitting experimental and
simulated curves allowed simultaneous determination of both the wall shear
rate and Hamaker constant. It may be emphasized that this method was highly
sensitive since it allowed to detect forces of order of 10 femtonewtons.
Fitted parameters were used to plot the binding frequency of flowing spheres versus
dimensionless ratio $U/aG$,  where $U$ is the sphere average velocity over a 
160 ms interval, $a$ is the sphere radius, and $G$ is the wall shear rate.

 In contrast
with standard results from fluid mechanics\cite{Goldman} , $U/aG$ could only be
related to a distribution probability of sphere-to-surface distances. It was thus
not feasible to derive analytically the distance-dependent association rate $k_{on}(d)$ 
from the binding curve. Thus, in order to proceed, simulated trajectories were built
by assuming a simple form for $k_{on}(d)$ : the rate of association between individual
molecules anchored at two points separated by a distance $d$ was taken as a constant
$k_{on}$ when $d$ was lower than a threshold range $R$ and zero elsewhere. Experimental
data were consistent with the conclusion that $k_{on}$ was about $1.2\ 10^{-3} s^{-1}$ and $R$ 
was about 10 nm. It may be tempting to compare theses estimates to the 3-dimensional
association rates : for this purpose, it may be noticed that the concentration of a
molecule restricted to a sphere of 10 nm radius is about 0.0004 M, yielding a tentative
three-dimensional $k_{on}$ of 3 $M^{-1}s^{-1}$. This is much lower than expected
association rates of soluble molecules. Two possible explanations may be considered~:

i) geometric restreints due to rigid coupling between molecules and surfaces may drastically
alter binding efficiency.

ii) The concept of association rate should be reconsidered when contact times as low
as 1 ms are considered : indeed, if a minimum contact time is required to allow sufficient
bond strenghtening that cadherin-cadherin association be detectable, the probability
of bond formation during a cell-to-surface approach my not be proportional to contact time.
This would make the concept of $k_{on}$ questionable.

\subsubsection{Comparison with other methods}
Atomic force microscopy was very early used to estimate intermolecular association
 rates\cite{Hinterdorfer96,Baumgartner00}. The basic idea\cite{Hinterdorfer96} was to
combine i) estimate of the time required for binding by varying the contact time on a
given position, and ii) estimate of the interaction range by combining very slow lateral
displacement and continuous binding-unbinding. Studying the interaction between human
albumin and polyclonal antibodies, the association rate was estimated at 
$5\ 10^4 M^{-1} s^{-1}$ which was deemed reasonable. In a later study, the association
rate for VE-cadherin was estimated between $10^3$ and $10^4 M^{-1} s^{-1}$. In line with 
previous remarks, it may be noticed that association rates might appear higher than 
estimated for the flow chamber since i) the association time was determined in 
regions were binding molecules were found. This removed the possibility that inactivity
of a fraction of receptors might result in too low values of association rates, and 
ii) contact times were on the order of 0.1 s, i.e. at least tenfold higher than used
in the flow chamber.

\subsection{Influence of binding site environment of association properties}
\subsubsection{Results obtained with the flow chamber}
While most studies devoted to ligand-receptor interaction are focussed on the relationship
between the molecular properties of binding sites and parameters such as association
and dissociation rates or mechanical properties, much experimental evidence supports the
view that receptor efficiency is higly dependent on a number of structural features
that are independent of binding sites, including mode of anchoring, length and 
flexibility of the connection with cell surfaces, as well as surrounding molecules.

A prominent example is the glycocalyx influence on membrane receptor efficiency.
 This was demonstrated by
elegant experiences related to P-selectin, an endothelial cell receptor interacting
with PSGL-1, a mucin ligand expressed by circulating leukocytes\cite{Patel95}. Both
P-selectin and PSGL-1 are large molecules of more than 40 nm length. It was deemed interesting
to study the binding capacity of engineered P-selectin with shortened stalk and intact
binding sites. Under static conditions, transfected cells expressing wild-type or 
mutated selectins bound leukocytes with comparable efficiency. However, when experiments
were conducted under dynamic conditions in a flow chamber operated at a wall shear rate
of about 200 $s^{-1}$, shortened molecules displayed drastically decreased capacity to
tether moving neutrophils. Further, binding capacity under flow was somewhat restored when
shortened P-selectin molecules were borne by defective cells with reduced glycocalyx
expression.

Similar conclusions were obtained with a flow chamber operated at a much lower shear 
rate of 10-20 s$^{-1}$. Phagocyte monolayers were allowed to bind to microbeads
interacting with immunoglobulin receptors expressed on cell membranes. Under static 
conditions, binding was comparable in control cells and cells that had been treated
with neuraminidase to reduce their glycocalyx. On the contrary, glycocalyx reduction
drastically enhanced microbead uptake under flow\cite{Sabri95}.

\subsubsection{Other methods aimed at studying individual molecular bonds formed between 
surface-attached molecules support the importance of binding site environment on 
bond properties}
The importance of binding site environment was repeadedly revealed in a number of studies.
Thus, initial studies made on single bond formation with atomic force microscopy
stressed the advantage of depositing a molecular species on soft agar beads\cite{Florin94}.
Using a dual micropipette system to determine the frequency of bond formation between 
soft vesicles subjected to transient contact, it was reported that the frequency of
bond formation, not dissociation, between ligand and receptor molecules was decreased
when molecular length was reduced\cite{Huang04}. Also, the surface forces apparatus was
used to demonstrate that the efficiency of association between tethered streptavidin and
biotin molecules was increased in proportion with tether length\cite{Jeppesen01}.

\subsection{Studying the behaviour of attachements mediated by multiple bonds}
Under most conditions, adhesion is mediated by multiple bonds. Therefore, there
is a strong need to relate single and multiple bond properties. It is difficult
to achieve this goal by purely theoretical means. Indeed, conclusions are heavily
dependent on assumptions that are difficult to assess. Thus, in a recent analysis
of the rupture of multiple parallel bond, rupture force was concluded to vary
according to the first power, the square root or the logarithm of bond number depending
on parameters such as force sharing between multiple bonds or possibility of 
rebinding\cite{Seifert2000}. Experimental studies are thus needed.

\subsubsection{Quantification of the properties of multivalent attachment with
 a flow chamber}
There are many examples of cell surface receptors that seem to require a minimal 
oligomerization for full binding activity. Thus, ICAM-1, the ubiquitous ligand
of leukocyte integrin LFA-1, has long been reported to be a dimer on cell membranes.
Recently, it was shown that soluble ICAM-1 monomer bound LFA-1 with maximal affinity, but 
dimerizaton was required to mediate efficient attachment between membrane-bound ICAM-1
and LFA-1\cite{Jun01}. Also, molecular clustering has long been considered as a general
mechanism of integrin activation. This was an incentive to quantify the effect of 
clustering integrin $\alpha 5\beta 1$ on binding efficiency. The integrin-mediated
attachment of monocytic THP-1 cells to fibronectin-coated surfaces was studied with a
laminar flow chamber operated at low shear rate\cite{Vitte04}. Results are summarized
on Table~4 :

\begin{center}
{\bfseries Table~4. Effect of integrin aggregation  on binding efficiency under flow.}
\begin{tabular}{lccc} \hline
Cell treatment &Fibronectin concentration (mol $\mu m^{-2}$&Binding frequency
$mm^{-1}$&Initial detachment rate $s^{-1}$\\
\hline

None & 6,500 & $1.48\pm 0.07$ & $0.96\pm 0.10$\\
anti-integrin mAb & 6,500 & $0.74\pm 0.07$ & $1.64\pm 0.26$\\
Anti-integrin+anti-Ig& 6,500 & $0.93 \pm 0.08$ & $1.18\pm 0.20$\\
\\
None & 3,850 & $0.75\pm 0.08$ & $2.26\pm 0.40$\\
anti-integrin mAb & 3,850 & $0.45\pm 0.04$ & $1.83\pm 0.29$\\
anti-integrin+anti-Ig& 3,850 & $1.19\pm 0.08$ & $1.02\pm 0.14$\\
\\
None & 1,400 & $0.21\pm 0.02$ & $1.94\pm 0.27$\\

\hline \end{tabular} 
\end{center}

First, the effect of fibronectin surface density on binding frequency and duration
was studied on control cells. When fibronectin surface density was decreased from
about 6,500 to 3,850 molecules/$\mu m^2$, both binding frequency and arrest lifetime
were decreased, suggesting that attachments might be multivalent. However, when 
fibronectin surface density was further decreased, binding frequency was strongly
decreased without any significant alteration of bond lifetime, which suggested
that individual bonds were indeed observed.

Now, when cells were treated with a supposedly neutral anti-integrin antibody, binding
frequency on high or intermediated density fibronectin was lowered, and detachment
rates were comparable to values ascribed to single bond. The simplest interpretation
was that antibodies decreased integrin accessibility, not function.

Finally, when a second (anti-immunoglobulin) antibody was added in order to mediate
receptor aggregation (which was checked with confocal microscopy), binding
frequency was increased.

A simple interpretation would be that receptor aggregation might favor multivalent
initial interactions. This would increase both binding frequency and bond lifetime
if it is assumed that binding events detected with a flow chamber might result from
the strengthening of shorter - undetectable -binding states. While further work is
clearly needed to understand quantitatively the mechanisms of initial adhesion, 
presented data clearly support the view that cell adhesive behavior may depend
on fine properties of the entire energy landscape of ligand-receptor complexes.

\subsubsection{Studying multivalent attachments with other methods}
The diversity of results described in different reports supports our expectation that
there is probably no single rule to predict the behaviour of multivalent attachments.
Thus, in an early study of the rupture of streptavidin-biotin bonds with atomic
force microscopy\cite{Florin94}, the histogram of rupture force frequency displayed
a series of peaks appearing as integer multiples of an elementary force quantum
(namely 160 pN) attributed to a single bond, thus suggesting simultaneous breakage
of all bonds). In another study made on antigen-antibody interaction\cite{Hinterdorfer96},
the bonds formed by a single divalent antibody appeared as separate events. In 
a recent report\cite{Ratto06}, the rupture of multivalent attachments between
concanavalin A, a protein bearing four carbohydrate binding sites, and mannose
groups suggested nonlinearly additive forces since the rupture forces of attachments
involving 1, 2 and 3 bonds respectively were respectively 46, 68 and 85 pN.

\section{Significance of experimental data and problems raised by flow chamber experiments}

Although flow chambers have now been used for more that ten years to study molecular 
interactions, some important points still need clarification. We shall now address
problems that seem to deserve some attention. Additional information may be
found in a recent review\cite{Zhu2002}.

\subsection{Is there an unambiguous means of proving that single molecular bonds are
actually observed ?}

We shall now briefly list some arguments that were sometimes used.

{\it When bond formation is rare, multiple bond formation becomes negligible.}
This argument was often used with other techniques such as atomic force microscopy.
Thus, if only 10 \% of approach/retraction cycles are conducive to bond formation,
it is often concluded that only 1 \% of these cycles will be conducive to multiple
bonds. However, this argument is not tenable unless it is assumed that all binding
events mediated by single bonds are actually detectable. Otherwise, it may only be 
concluded that single {\it minimal detectable events} are studied.

\begin{figure} [ht] \begin{center}
\includegraphics[width=10cm]{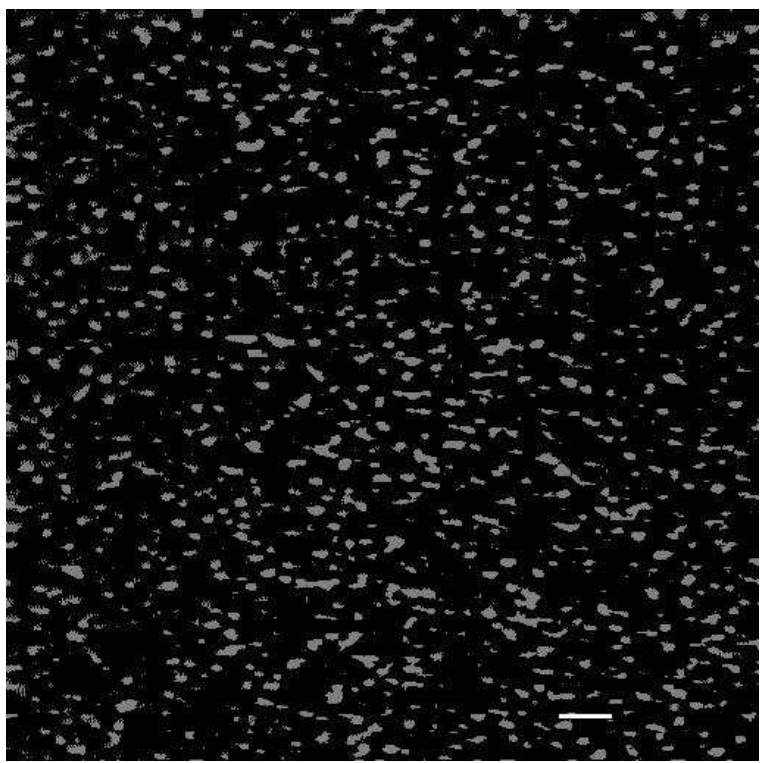}
\vspace*{8pt}
\caption{\small A sample image of a mica surface coated with fibronectin at about
1,400 molecules/$\mu m^2$. The image was obtained with atomic force
microscopy in aqueous medium and tentatively binarized with a height
threshold of 9.2 nm. This may resemble what is actually seen by
receptors on the surface of flowing cells. Clearly, fibronectin is 
not organized as an assembly of isolated molecule with identical conformation.
 This illustrates the need
 of studying the molecular
organization of adhesive surfaces. Bar length is 250 nm.}
\label{figafm} \end{center}
\end{figure}

{\it When binding sites are diluted, the frequency of binding events decreases
linearly with respect to binding site density.} This argument may not be considered
as a fully rigorous proof that single bonds are observed. Indeed, if a proportion 
of receptors or molecules are aggregated, which is a fairly frequent occurrence
with protein samples, it is conceivable that multivalent bond frequency might
behave as a linear function of receptor density. Note that this possibility is 
difficult to rule out since ligand adsorption on a surface may be somewhat irregular
even if coating media were carefully deaggregated, e.g. by ultracentrifugation.
A sample image of adsorbed fibronectin is shown as an example on Fig.\ref{figafm}.
 
 A possible way of ruling out this
possibility might consist of comparing the frequency of binding events to the 
expected encounter-frequency of ligand and receptor sites. Clearly, if a high
proportion of these encounters resulted in detectable binding, these events should
not represent the presence of a rare proportion of aggregates.

{\it Unbinding events follow first order kinetics.} As was previously 
emphasized\cite{CellMechanics94}, if the minimal duration of detectable binding
events is too high, the rupture of multivalent attachments may
seem to follow first order kinetics if the initial nonlinear part of the
curve is missed.

{\it When binding sites are diluted, binding event frequency decreases while 
the unbinding kinetics remains unchanged.} Obviously, this argument may rule
out the possibility that observed binding events are a combination of unimolecular
and multimolecular associations. However, this is fully convincing only if it is
shown than minimal detectable events are single molecular bonds.

{\bf Conclusion}. While it seems reasonable to conclude that many authors
actually monitored individual bond formation and dissociation, much caution
is needed to prove that this happened in a given experimental situation.

\subsection{Is it possible to quantify the force dependence of unbinding
rates with a flow chamber ?}

Flow chambers were repeatedly used to determine the mechanical properties
of single bonds\cite{Alon95,Pierres1996,Puri,Perret02}. The basic principle
consisted of determining bond dissociation rates for different values of
the wall shear stress. The force applied on the bond was derived from
the force on the cell with simple geometrical arguments\cite{Alon95,Pierres1995}.
This procedure would be fully warranted if bond formation was always
a monophasic process. However, when bond formation is multiphasic, a
decrease of bond lifetime as a consequence of an increase of the wall shear
rate may be ascribed to either a decrease of bond lifetime in presence
of a disruptive force, or a decrease of bond strengthening due to decrease
of the time allowed for bond formation before being subjected to hydrodynamic
drag. As was mentioned above, unperturbed intermolecular contact may last
a millisecond or less in a flow chamber, and there is much evidence to support
the view that formation of a fully stable bond may require a notably higher
amount of time\cite{Pierres1995,Pierres2002,Pincet2005}. It is therefore
conceivable that the use of flow chambers with too high shear rates might
result in the observation of multiple incompletely strengthened bonds,
thus accounting for possible discrepancies with force parameters obtained
with other methods.

\subsection{What is the meaning of binding frequencies determined with flow
chambers ?}
A notable advantage of operating flow chambers under fairly low shear rate is
that cell or particle trajectories may be easily followed independently of
the occurrence of any binding event. Thus, it is easy to determine the
binding probability per unit length of trajectory. Now, there remains to assess
the significance of this parameter. Two different hypotheses may be
 considered.

 i) The rate limiting step might be the molecular encounter between receptors
 and ligands. If each encounter is conducive to bond formation, binding
 frequency per unit length should be independent of particle velocity.

 ii) Alternatively, if the intrinsic rate of bond formation is low, contact
 time might be the rate limiting step. Increasing particle velocity would
 then be expected to decrease binding frequency.

 Interestingly, recent experimental evidence suggests that both possibilities
 may occur\cite{Vitte04JLB}. Indeed, when human lymphocytic Jurkat cells were
 made to bind to ICAM-1, a ligand of cell surface integrin LFA-1, binding
 frequency was inversely correlated to the wall shear rate. However, when
 the flow chamber surface was coated with anti-LFA-1 antibodies, binding
 frequency per unit length became independent of the wall shear rate. This
 difference is consistent with the notoriously high association rate of
 antibodies.

 {\bf Conclusion}. Before trying to extract information on molecular association
 rates from binding frequency measurements, it is important to check that
 adequate experimental conditions (i.e. ligand density and wall shear rates)
 have been selected.

 \subsection{Is molecular on-rate an intrinsic parameter ?}
 A rather subtle cause of misinterpretation consists of trying to determine
 a parameter that does not actually exist. A prominent example is the
 study of bond rupture strength with atomic force microscopy. Initial
 results were interpreted with the underlying assumption that there exists
  an intrinsic bond strength, which is true only when temperature is low
  enough. Similarly, $k_{on}$ determinations rely on the assumption that
  the probability that a ligand and a receptor form a bond during an encounter
  of vanishingly short duration $dt$ is proportional to $dt$. In fact, this
  procedure raises some difficulties. First, defining a ligand as bound to
  a receptor requires the determination of an arbitrary threshold, be it
  a minimal binding energy or capacity to resist a disruptive force during
  a sufficient period of time to be detected. Second, several barriers may
  have to be crossed for such a threshold to be reached. In this case, the
  binding probability may be proportional to a power of encounter duration
  that might be higher than one. An other way of describing this problem
  would be to recall that a rigorous study of association rate would consist
 of i) defining binding states, and ii) determining the probability $k_{on}(d,t)$
 that a ligand and a receptor molecules remained "bound" after remaining
 separated by distance $d$ during time $t$.

 \subsection{The limits of the flow chamber method are set by the requirement
 to discriminate between actual and artefactual binding events}
 First, it must be emphasized that the limits of bond analysis may be set
 by qualitatively different parameters when cells or microspheres are
 used. Indeed, an essential feature of biological cells is that they are
 studded with protrusions of varying size, and they may resemble ellipsoids
 rather than perfect spheres. This results in two difficulties : first, the
 region of contact between a cell and the chamber floor under flow is not
accurately known. Thus, if a cell is rotating around a bound contact, a
bond may remain undetectable since the cell may seem to move continuously.
Conversely, a cell may seem to undergo a brief arrest if it rotates on the
tip of a large protrusion\cite{Tissot92}. Thus, it was felt that it was
better to use microspheres rather than cell to achieve optimal resolution
in studying molecular interactions\cite{Pierres1995,Pierres1996}. Indeed,
The accurate position of sphere-to-surface contact may be determined with
high accuracy by calculating the position of the centroid of the sphere
image. It was thus estimated that 50 nm accuracy was easily obtained 
when spheres of 2.8 $\mu m$ diameter were monitored. Now, three different
artifactual binding events might occur. These will be considered sequentially.
For this purpose, the simplest procedure consists of defining a sphere as
arrested when it moved by a distance lower than some threshold $\chi$ during
a given period of time $\tau$

\begin{figure}[ht] \begin{center}
\includegraphics[width=10cm]{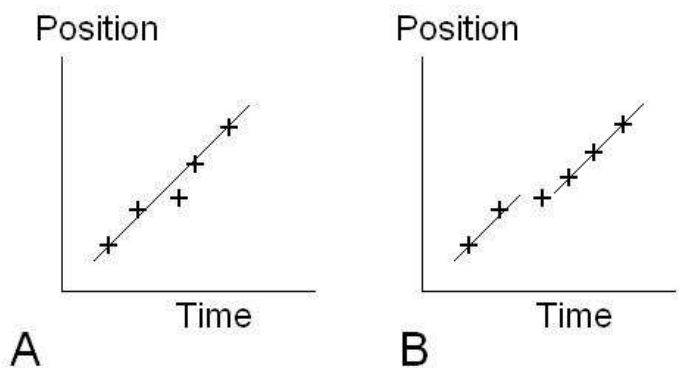}
\vspace*{8pt}
\caption{\small Errors in position determination. Artefactual arrests due to 
errors in position determination are easily detected on visual inspection
(A) and they appear as isolated points. Actual decrease of particle
velocity result in translation of the regression line built on a sequenc
of steps (B).\label{ErArret}}
\end{center} \end{figure}

{\it Consequences of measurement errors}.
Clearly as shown on Figure \ref{ErArret}, an error of position determination
may result in artefactual short arrest of about 1 step duration. This is
fairly easy to detect on visual examination of the trajectory, and this error
might be detected by using a more refined way of defining arrest, such as
a translation of the regression line determined on a series of steps
\footnote{PB is indebted to Dr. D. Bensimon for pointing out the potential
interest of refining arrest detection}. However, we did not find it warranted
since it was only a minor problem as compared to other artifacts. Note that
it is essential to rule out more important problems due to collisions
between flowing and arrested particles. This is easily controlled by
combining position and area determination, and deleting events associated
with transient area increase\cite{Pierres97}

{\it Statistical errors due to Brownian motion}. 
In a flow chamber, Brownian motion will result in displacements parallel to
the flow (i.e. horizontal) and perpendicular to the wall (i.e. vertical).
The latter displacement will generate a velocity variation proportional to
the wall shear rate G, since the average velocity of a sphere is dependent
on the distance to the wall\cite{Goldman}. The resulting variance of the
displacement $\Delta x$ during time interval $\Delta t$ is thus expected to
follow the equation :

\begin{equation}
<\Delta x>^2 = a \; G^2 <\Delta t>^2 +\; b \;  \Delta t
\label{vardx} \end{equation}

An important consequence is that the hydrodynamic displacement, which is
proportional to $<\Delta t>$ will decrease more rapidly than the Brownian
displacement when the time period vanishes. This means that it is useless to
increase the sampling rate for position determination. In other words,the
time resolution limit for arrest detection is set by Brownian motion.
Note that the validity of Equation \ref{vardx} was checked both experimentally,
by studying the motion of non adhesive particles, and with computer
simulation. Now, for a given wall shear rate, the threshold parameters
for arrest detection must be chosen in order to maximize detection
sensitivity {\it provided the proportion of statistical binding events remain
negligible as compared to physical binding events}. This means that
arrest detection may be more sensitive when efficient binding molecular
models are studied. On the contrary, when physical binding is unfrequent
and many positions must be studied, higher threshold time $\tau$ must be
selected. Thus, if 100,000 displacement steps must be scanned to detect
about 100 binding events, statistical artefact probability must be lower than
$10^{-4}$ in order that 90 \% of detected arrests represent {\it bona fide}
binding events.
A possible way of reducing the importance of statistical arrests might
seem to consist of increasing the wall shear rate. However, this will often
reduce binding frequency. This means that the optimal conditions for
studying a given molecular model must be determined somewhat empirically.
A parameter that might seem worth considering is particle size.
However, reducing particle size in order to improve the ratio between
sphere hydrodynamic velocity and force on a bond will increase vertical
brownian motion, thus decreasing the sphere residence time within binding distance
of ligands. Binding efficiency might thus be drastically decreased.

\subsection{Treatment of nonspecific interactions}
A problem that is always encountered when starting to study a new molecular model
is the following : how can we prove that monitored binding events are {\it bona
fide} interactions between ligand and receptor binding sites ?

An apparently reasonable check might consist of verifying that particles no longer
bind the chamber floor when either particle or chamber coating is omitted. However,
this is not a rigorous proof. Thus, that bare spheres do not bind to a ligand-coated
surfaces is not a proof that receptor-coated spheres will form specific interactions
with the same surface. Indeed, receptor molecules might form nonspecific interactions
with the chamber surface.

Another argument might consist of checking that binding events are inhibited when
the flow medium is altered in order to prevent specific ligand-receptor bonds. A
frequent way of achieving this results might consist of adding divalent cation chelators
since these cations are required for binding mediated by important adhesion receptors
such as selectins, integrins or cadherins. Unfortunately, performing this test cannot
be considered as a satisfactory check. Thus, in our laboratory, cadherin coated spheres
displayed calcium-dependent binding to cadherin-coated surfaces, as expected. However,
calcium chelation also prevented association between cadherin-coated beads and bare
mica. In other words, nonspecific interactions may display a similar sensitivity to 
physico-chemical conditions as specific associations do.

Also, according to our experience, nonspecific interactions do not exhibit clearcut 
enough properties such as hight sensitivity ot rupture forces, to allow safe discrimination
between the properties of specific and nonspecific bonds.

Thus, according to our experience, the safest way of proving that specific ligand-receptor
bonds are indeed studied may consist of specifically inhibiting binding events with
small molecular weight molecules or comparing wild-type receptors to genetically engineered
molecules wiht a critical mutation located to the binding site. Thus, streptativin-biotin
interactions were efficiently inhibited by adding low molecular weight biotin to the 
surrounding medium\cite{Pierres2002}. Also, homophilic cadherin interactions were inhibited
by mutating of a critical tryptophan residue into alanin\cite{Perret02}.

\subsection{Intrinsic ambiguity of data interpretation}
A significant difficulty arose as the analysis of single molecular bonds was more and more
refined. Several different models can account for a given set of experimental data. This 
intrinsic ambiguity was indeed quoted in several experimental\cite{Evans04} or
 theoretical\cite{Derenyi04} studies. The analysis of unbinding curves obtained with flow
chambers clearly illustrates this point : a typical curve such as shown on Fig. \ref{f2} appears
with upwards concavity that might be due to i) formation of additional bonds after initial
arrest\cite{Pierres1996}, ii) time-dependent strengthening of a single bond\cite{Pierres1995} or
iii) existence of two different bond species with different rupture frequencies. It is possible to
discriminate between (i) and (ii) by performing different experiments with varying ligand
or receptor surface concentrations. However, (ii) and (iii) may be difficult to distinguish
if it is accepted that a given ligand-receptor pair may form two different kinds of bonds, as
was recently demonstrated\cite{Sivasankar01}. Indeed, when ligand-receptor interaction consists
of forming a transient state with dissociation rate $k_{off}$ and rate of transition to a 
stable state $k_t$, the equation of the unbinding curve is readily shown to be~:

\begin{equation}
N(t) = \frac {N_0 \; [k_m \; \exp(-(k_{off} + k_t)t) \; + \; k_t]}{k_{off} + k_t}
\label{ModTrans} \end{equation}

Howevever, the unbinding rate of particles that may be bound either with a transient bond with
dissociation rate $k_{off}$ or a stable bond is given by :

\begin{equation}
N(t) = A \; \exp(-k_{off}t) \; + \; B
\label {Double} \end{equation}

Clearly, Equations \ref{ModTrans} and \ref{Double} are formally identical.

 Thus, more and more complementary experiments will be
needed to achieve safe interpretation of experimental data.

\subsection{Accuracy of rate constant determination.}
An important point to achieve meaningful comparizon of results obtained with different techniques
is te determine the accuracy of rate constant determination. A prominent cause of variation is 
due to counting errors. This may be illustrated by considering off-rate determinations. Computer
simulations were performed to determine the minimum number of arrests used to build unbinding
curves\cite{MassonGadais99}. It was concluded that the coefficient of variation of the off-rate
determination was respectively 39\%, 25\%, 17\% and 8.5\% when 25, 50, 100 or 200 arrest durations
were used.

Another simple estimate was obtained as follows\cite{Pierres2002} : assuming Poisson distribution
for the numbers A and B of particles detached at time t within interval $[t_1,t_2]$ and
$[t_2,\infty]$ after arrest, the average detachment rate during interval $[t_1,t_2]$ is
$\frac{\ln(B/(A+B))}{t_2 \; - \; t_1}$. The standard deviation for dissociation rate is therefore
equal to $\sqrt{\left[ \frac{A}{B(A+B)} \right] }/(t_2 \;- \; t_1)$

\section{Conclusion.}
The aim of this brief review was essentially to summarize the new kind of information that was 
made available by using a flow chamber operated under low shear rate to dissect the interaction
between surface-attached ligand and receptor molecules at the single bond level. The following
conclusions were obtained.
i) The rates of bond formation and dissociation between soluble ligand and receptors do not
encompass sufficient information to account for the whole functional activity of these molecules
when they are bound to cell surfaces. Indeed, parameters such as molecular length, attachment 
flexibility, topographical distribution on the cell surface and molecular environment strongly
influence adhesive interactions.
ii) The flow chamber was found to allow efficient analysis of the functions of most adhesion
molecules such as selectins\cite{Kaplanski93,Alon95}, 
integrins\cite{MassonGadais99,Vitte04,Vitte04JLB}, members of the immunoglobulin
 superfamily\cite{Pierres1996} or cadherins\cite{Perret02}.
iii) Results obtained with this and other techniques suggest that the rupture of newly
formed attachments may often be adequately described with a set of {\bf three} parameters, i.e.
rupture frequency and force coefficient of initial binding state and strengthening rate.
More work is required to find an adequate set of parameters to describe bond formation.
iv) In view of presently available evidence, it is certainly warranted to undergo systematic
comparison between bond parameters obtained with different methods, including flow chambers,
atomic force microscopy and biomembrane force probes, and then to determine whether these
parameters give a satisfactory account of receptor function under physiological conditions.
v) Another important prospect is certainly to pursue comparisons between experimental 
parameters and computer simulation of bond formation and dissociation\cite{Bayas03}. This
is certainly a promising way of gaining more profound understanding of protein structure and
function, as well as predicting functional properties of receptors of known structure.
vi) However, it must be kept in mind that data yielded by flow chambers are often 
difficult to understand completely, and full interpretation often requires a number of
controls that must be specifically to each system under study.

\section*{Acknowledgments}
Part of presented work was supported by the ARC.

\section{references}

\end{document}